\documentclass[aps,prl,superscriptaddress,twocolumn,showpacs,amsmath,amsfonts,amssymb,floatfix]{revtex4}
\usepackage{graphicx}
\usepackage{epsfig}
\begin{document}
\title{Circular hydraulic jump in generalized-Newtonian fluids}

\author{Ashutosh Rai}\email{arai@iitk.ac.in}
 	\affiliation{Indian Institute of Science Education and Research, Salt Lake, Kolkata - 700106, India.}
\author{B.S. Dandapat}
 	\affiliation{Physics and Applied Mathematics Unit, Indian Statistical Institute, B.T. Road, Kolkata - 700108, India.}
\author{Swarup Poria}
    \affiliation{Dept. of Mathematics, Midnapore College, Midnapore - 721101, India.}

\begin{abstract}
We carry out an analytical study of laminar circular hydraulic jumps,
in generalized-Newtonian fluids obeying the two-parametric
power-law model of Ostwald-de Waele. Under the boundary-layer approximation
we obtained exact expressions determining the flow and an implicit relation for
the jump radius is derived. Corresponding results for Newtonian fluids can be
retrieved as a limiting case for the flow behavior index $n=1$, predictions are
made for fluids deviating from Newtonian behavior.
\end{abstract}


\maketitle

When a smooth jet of liquid falls vertically on a rigid horizontal boundary,
it spreads out radially in a thin layer until a sudden increase
in depth (of a circular geometry) may occur; this is a \emph{Circular Hydraulic Jump} (CHJ) (see Fig.1).
CHJ is an observable in everyday life as one opens a water tap into a kitchen sink. Apart from this simple
day to day observation, striking analogies with CJHs have
been seen at femto-scale where a CHJ appear as metal femtoliter cups \cite{Mani}, and also at astronomical scales where it is manifested as
a white hole horizon \cite{Volo}.

In this letter we present new results on CHJs in generalized-Newtonian fluids, where,
the effective viscosity depends, in general, on shear rates of the flow, which leads us to interesting results applicable to
a larger class of fluids, results for Newtonian fluids can be obtained as a special case.
In contrast to our study, earlier studies on CHJs has been done only for the Newtonian fluids.

\begin{figure}
\includegraphics[scale=0.33]{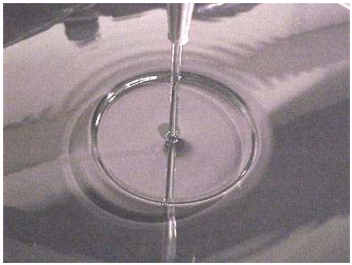}\caption{\label{FIG. 1} A circular hydraulic jump (CHJ) is formed when a jet of liquid
falls vertically on a rigid horizontal boundary (Photo courtesy of John Bush and Jeff Aristoff, MIT) }
\end{figure}

CHJs were noticed and theoretically investigated
for the first time by Rayleigh in 1914, who developed a theory consistent for inviscid flow
by conserving mass flux and momentum flux across the jump \cite{Rayleigh}. However, later studies
revealed that due to divergence in the flow, the flow layer becomes very thin with increase in radial distance,
where viscosity plays a significant role in the flow dynamics \cite{Watson,Olson,Craik,Bohr1,Ell1}.
Watson in 1964 considered the influence of fluid viscosity on CHJs \cite{Watson}. In the thin layer
of the flow, \emph{boundary-layer theory} was applied to model the flow \cite{Schlichting}.
At the jump, application of the momentum theorem gave a prediction for the jump radius.
Experiments of Watson \cite{Watson}, Olson and Turkdogan \cite{Olson}, Craik \emph{et al.} \cite{Craik},
and Liu and Lienhard \cite{Liu} characterized the domain in which
Watson's theory showed good agreement with experiments, at the same time the region where the theory seemed weaker was also marked.
The agreement was good for smaller jump height to jump radius ratio and
poor in the opposite limit of relatively, large jump heights or small jump radii. Origin of the discrepancies was found in
Watson's simplifying assumptions: unidirectional flow beyond the jump and neglecting surface tension effect.

Several experimental and theoretical studies since then have revealed that CHJs have broadly two types of steady states,
classified as type I and type II \cite{Bush,Ell2,Yokoi}.
In experiments, the depth $H$ beyond the jump is controlled by a circular boundary of variable height $H$,
sufficiently far from the jump. Type I steady state is observed below a certain critical value $H_{c}$,
beyond which transition to type II steady state occurs.
In type I steady state, where surface flow is unidirectional and boundary-layer separation occurs beyond the jump,
Watson's assumptions were found to be appropriate. On the other hand type II steady state is
distinguished by reverse surface flow (roller) adjoining the jump, the boundary layer separation is pronounced
at the jump as a result of which the flow need not decelerate significantly as it passes through the jump:
the discontinuity in the radial speed assumed in Watson's work need not arise.

Steady jumps of type II is an active area of research till date, Bohr \emph{et al} \cite{Bohr1,Bohr2},
Higuera \cite{Hig}, Yokoi \cite{Yokoi} studied the boundary layer separation beyond the jump,
and developed explicit scaling law for the jump radius which conforms well with the experiments.
The key feature of these studies has been analysis of the complex flow dynamics near the jump, and its effect on the jump radius.
In contrast to above approaches, Watson's theory gives an implicit relation for the jump radius considering a
simplified average behavior of the flow just after the jump. Naturally as a consequence of this simplification, Watson's theory has
limitations, typically when the boundary-layer separation is dominant near the jump viz. steady type II jumps.
However, experiments suggest that in all types of CHJs: type I; type II or unsteady, Watson's
prediction for the jump radius provides an adequate leading-order description.
In the framework of Watson's theory, in 2003 Bush and Aristoff considered the
surface tension effect in Type I jumps, which led them to a revised estimate for
the jump radius \cite{Bush}. It was shown that in type I steady jumps of small radii and heights, surface tension plays
a significant role in the dynamical balance at the jump. An expression was derived for the radial force per unit length
associated with the curvature at the jump and Watson's result for the jump radius was appropriately revised.
Experiments done by the same authors \cite{Bush} showed good overall agreement with the theory for type I steady jumps, in particular
experimental data for jumps of small radius and height, which were poorly described by Watson's result, matched well on
including the surface tension correction.

CHJs so far has been studied (both theoretically and experimentally), only for the Newtonian fluids,
characterized by a viscosity coefficient, which does not depend on the shear rate of the flow.
Many fluids in nature e.g., oil, honey, paints, polymeric solutions, lime-water mixture etc,
are non-Newtonian, having an effective viscosity, which depends on the shear rates of the flows.
Boundary-layer theory in non-Newtonian fluids has been extensively studied by James and Dabrowaski, Filipussi \emph{et al.}
\cite{James}.
In this paper, we have modeled and studied the CHJ in generalized-Newtonian fluids
(a class of non-Newtonian fluids) obeying
two-parameter power-law model of Ostwald-de Waele \cite{Bird}. In this model, for one dimensional flows, shear stress,
$\tau$ can be given by $\tau = K\left(\frac{\partial u}{\partial z}\right)^{n}$,
where parameters $K$ and $n$ are the flow consistency index and flow behavior index
respectively ($0< n <1$ for a shear-thinning or pseudo-plastic fluid, for Newtonian fluids
$n = 1$ and $n >1$ for a shear-thickening or dilatant fluid). $\frac{\partial u}{\partial z}$
is the shear rate or the velocity gradient perpendicular to the plane of shear.
The quantity $\nu_{eff} = K\left(\frac{\partial u}{\partial z}\right)^{n-1}$
represents an apparent or effective viscosity as a function of the shear rate.
Contrary to the infinite range of
effective viscosity suggested by the mathematical form of the power-law model, in real fluids effective viscosity
is bounded for all possible shear rates.
Therefore, the power-law is approximate, and is only a good description of fluid behavior
across the range of shear rates to which coefficients are fitted.
Nevertheless power-law is the simplest and most widely used model to describe
the non-Newtonian fluid behavior, permit mathematical predictions,
and correlate experimental data.

Our study is in the framework of Watson's theory
in a sense that, assuming a continuous variation in flow behavior from a Newtonian to non-Newtonian fluid,
we have derived our results based on similar approximations as that of Watson \cite{Watson}. Laminar flow in thin layer
is studied under the boundary-layer approximation. Description of the flow is given in terms of Blasius sub layer
developing near the point of impact and a far-field similarity solution. The similarity-profile
of the far field similarity solution has been applied in the Blasius sub layer, along with the standard Karman-Pohlhausen approximation \cite{Schlichting}.
Momentum theorem is applied including the surface tension correction term
introduced by Bush and Aristoff \cite{Bush}, to obtain a prediction for radius of jump in power-law fluids.

On considering a smooth jet of fluid descending vertically on a horizontal plane
covered by the same fluid of depth $H$; from the point of impact the fluid spreads radially in a
thin layer, before the jump occurs. A boundary layer grows from the stagnation point up to a radial distance $r_{b}$
beyond which, it absorbs the whole of the flow. The jump occurs at some radial distance $r_{j}$.
Let, $a$ be the radius and $U_{0}$ be the velocity of the jet ( just before the impact),
$Q$ be the  volume flux of the flow, $U(r)$ the free surface stream velocity, and $h(r)$ height of free surface from the bottom,
$r$ being the radial distance from the point of impact (Fig. 2).
For $r \leq r_{b}$, the  flow speed outside the boundary-layer ( of thickness $\delta (r)$ )
remains almost constant, equal to $U_{0}$, as the fluid here is unaffected by the viscous stresses.
For $r \geq r_{b}$, the viscous stresses become appreciable right up to the free surface,
the whole flow is of the boundary-layer type.
\begin{figure}
\includegraphics[scale=0.4]{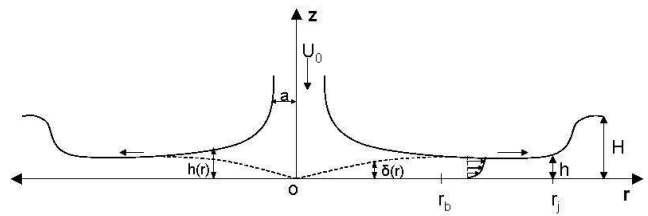}\caption{\label{FIG. 2} A schematic illustration of the boundary layer structure establised within the CHJ.
The viscous boundary layer grows until it reaches the surface at a radial distance $r_{b}$, from the point of impact,
$r_{j}$ is the jump radius.}
\end{figure}

For a stationary, radially symmetric flow with a free surface, boundary-layer equations for a power-law fluid are:
\begin{equation}
\label{eq1}
u \frac{\partial u}{\partial r}+w \frac{\partial u}{\partial z}=\frac{\partial }{\partial z} \Bigg(K \Bigg|\frac{\partial u}{\partial z}\Bigg|^{n-1}\frac{\partial u}{\partial z}\Bigg),
\end{equation}

\begin{equation}
\label{eq2}
\frac{\partial (ru)}{\partial r}+\frac{\partial (rw)}{\partial z}=0,
\end{equation}
where  $r$, $z$ are cylindrical co-ordinates, with $z$ measured vertically upwards from the plate, and $u(r,z)$, $w(r,z)$
are the corresponding velocity components. Gravitational pressure gradient has been neglected as in \cite{Watson}. The boundary conditions
are no-slip on the bottom: $u(r,0) = w(r,0)= 0$, no stress on the free surface: $\frac{\partial u}{\partial z}=0$, and the kinematic boundary
condition at the free surface: $w(h,r) = u(h,r)\frac{dh}{dr}$, which gives condition of constant volume flux
$Q = 2 \pi r \int^{h}_{0}u dz = \pi a^{2}U_{0} =$constant. Due to symmetry of the flow shear stress components $\tau_{rz}$ and $\tau_{z\theta}$
are zero, shear stress component $\tau_{r \theta} = K \Big|\frac{\partial (u)}{\partial z}\Big|^{n-1}\frac{\partial (u)}{\partial z}$.\\

In the region $r \geq r_{b}$, assuming far field solution of a similarity form: $u = U(r)f(\eta)$ where $\eta = \frac{z}{h(r)}$ , boundary conditions transform to:
$f(0)= 0$, $f(1)= 1$, $f'(1)= 0$, and condition of constant volume flux can be written as:
$Q=2\pi r Uh \int^{1}_{0}f(\eta)d\eta = $ constant.
On solving the equations of motion (\ref{eq1}) and (\ref{eq2}) for similarity solution, we found that similarity function is given by the implicit relation
\begin{equation}
\label{eq3}
\eta =\frac{1}{c}\int^{f}_{0} \frac{dx}{(1-x^{3})^{\frac{1}{1+n}}},
\end{equation}
where the constant: $c = \frac{1}{3}\beta(\frac{1}{3},\frac{n}{n+1})$. Here $\beta$ is the beta function and
$\int^{1}_{0}f(\eta)d\eta$ evaluates to $\frac{1}{3c}\beta(\frac{2}{3},\frac{n}{n+1})$ (say, $\equiv A$), free surface velocity $U(r)$
and free surface height $h(r)$ for $r\geq r_{b}$, turns out to be:
\begin{eqnarray}
\label{eq4}
 U(r)&=&\left[\frac{k(1-2n)}{n+2}\Bigg(\frac{2 \pi A}{Q}\Bigg)^{n+1}\hspace{-.5cm}(r^{n+2}+l^{n+2})\right]^{(\frac{1}{1-2n})}, \\
\label{eq5}
 h(r)&=&\frac{Q}{2\pi A r}\Bigg[\frac{k(1-2n)}{n+2}\Bigg(\frac{2 \pi A}{Q}\Bigg)^{n+1}\hspace{-.5cm}
(r^{n+2}+l^{n+2})\Bigg]^{-(\frac{1}{1-2n})}
\end{eqnarray}
the constant, $k = -3\left(\frac{n}{n+1}\right)Kc^{(n+1)}$ and $l$ is a length scale to be determined.
For the determination of length scale $l$, we proceed by considering the flow in the region $r\leq r_{b}$. The boundary-layer equation can be further
simplified by using a standard averaging technique of von Karman and Pohlhausen \cite{Schlichting}, we shall use the self-similar velocity
profile:  $u = U(r)f(\frac{z}{\delta})$, here $\delta$ is the boundary-layer thickness. As a result of averaging over $z$, the momentum
integral equation of the flow is:
\begin{equation}
\label{eq6}
\Bigg(\frac{d}{dr}+\frac{1}{r}\Bigg)\int^{\delta}_{0}(U_0u-u^2)dz=K \Bigg(\frac{\partial u}{\partial z}\Bigg)^n_{z=0},
\end{equation}
solving it for the self-similar velocity profile leads to,
\begin{equation}
\label{eq7}
\delta(r)=\left[\frac{B(n+1)}{(n+2)}r\right]^\frac{1}{n+1}.
\end{equation}
where the constant $ B=\frac{Kl U_0^{n-2}c^n}{A-\frac{n+1}{3nc}}$.
From here applying the condition of constant volume flux at $r = r_{b}$ gives,
\begin{equation}
\label{eq8}
r_b = \left[\frac{n+2}{B(n+1)}\Bigg(\frac{Q}{2\pi A U_0}\Bigg)^{n+1}\right]^\frac{1}{n+2},
\end{equation}
matching $U(r) = U_{0} $ at $r = r_{b}$ leads to:
\begin{equation}
\label{eq9}
l=\Bigg[\frac{\big[(1-2Ac)n+1+Ac \big](n+2)\pi^{n-2}}{(2n-1)(n+1)c^{n+1}(2A)^{n+1}}\Bigg]^{\frac{1}{n+2}}a \textit{Re}^{\frac{1}{n+2}},
\end{equation}
where, $Re = \frac{Q^{(2-n)}}{Ka^{4-3n}}$, is the jet \textit{Reynold's number}.
\begin{figure}[h!]
\begin{center}
$\begin{array}{c@{\hspace{.1in}}c@{\hspace{.1in}}c}
\includegraphics[width=8cm,height=5cm]{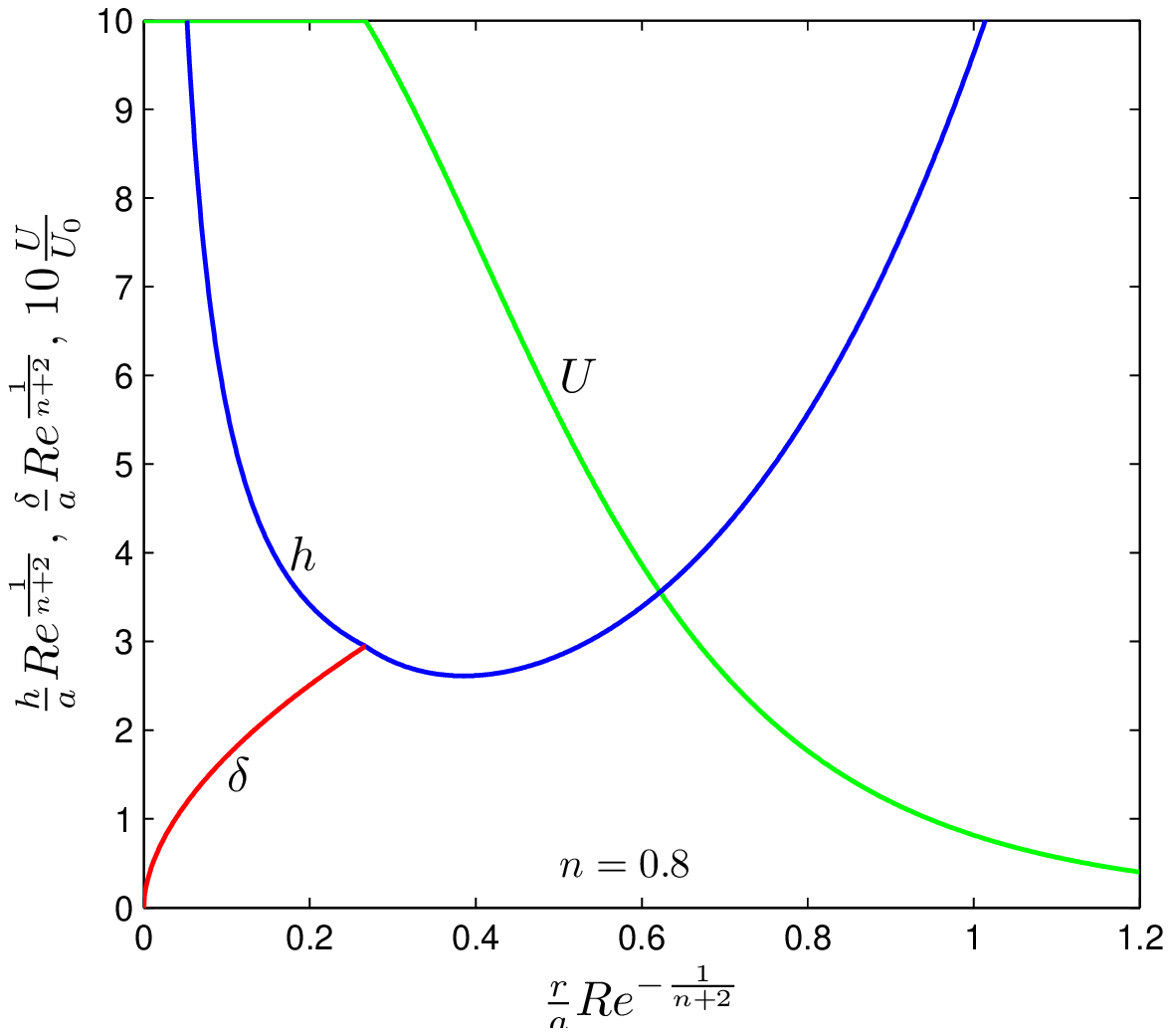}  \\
\includegraphics[width=8cm,height=5cm]{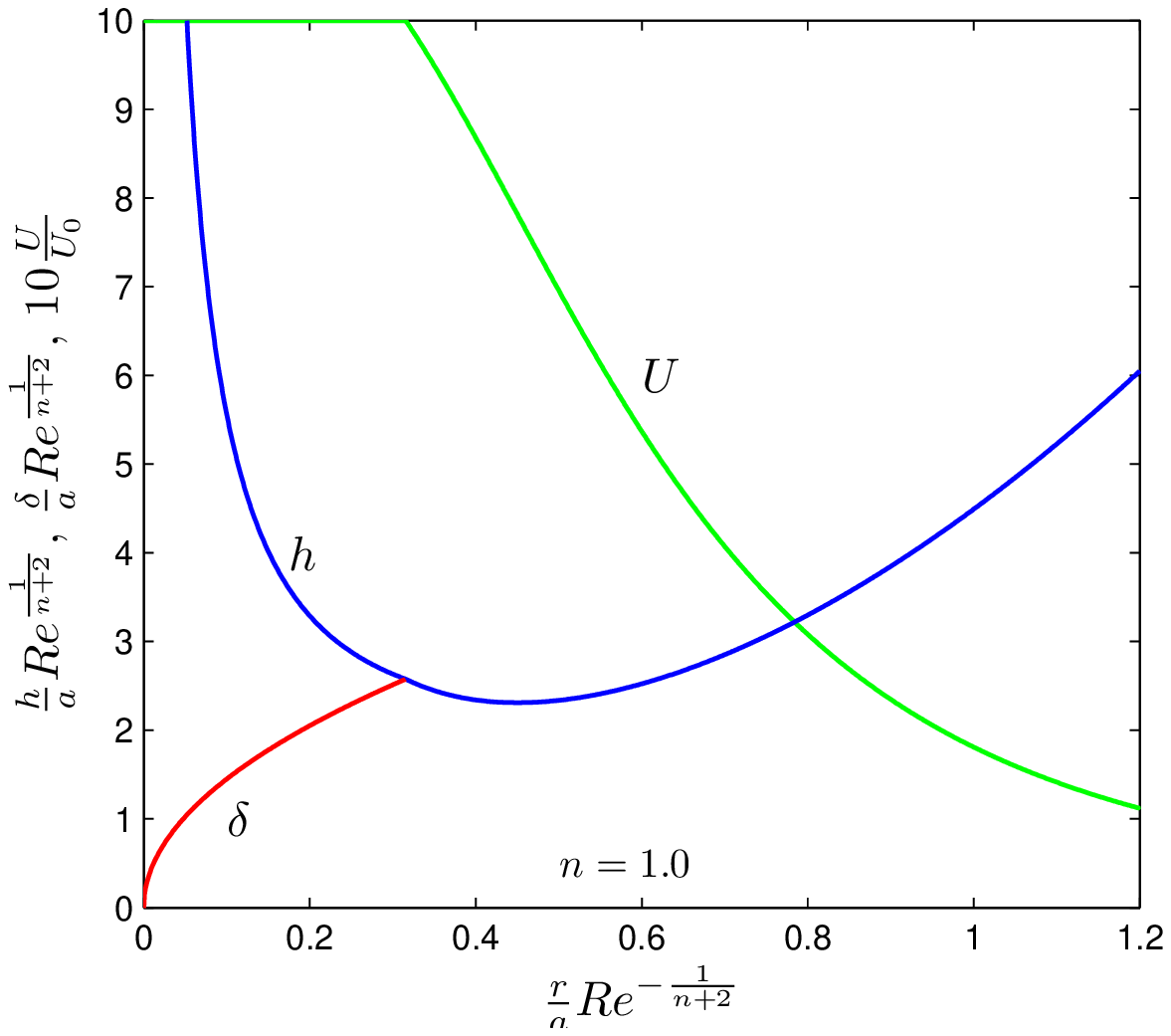}  \\
\includegraphics[width=8cm,height=5cm]{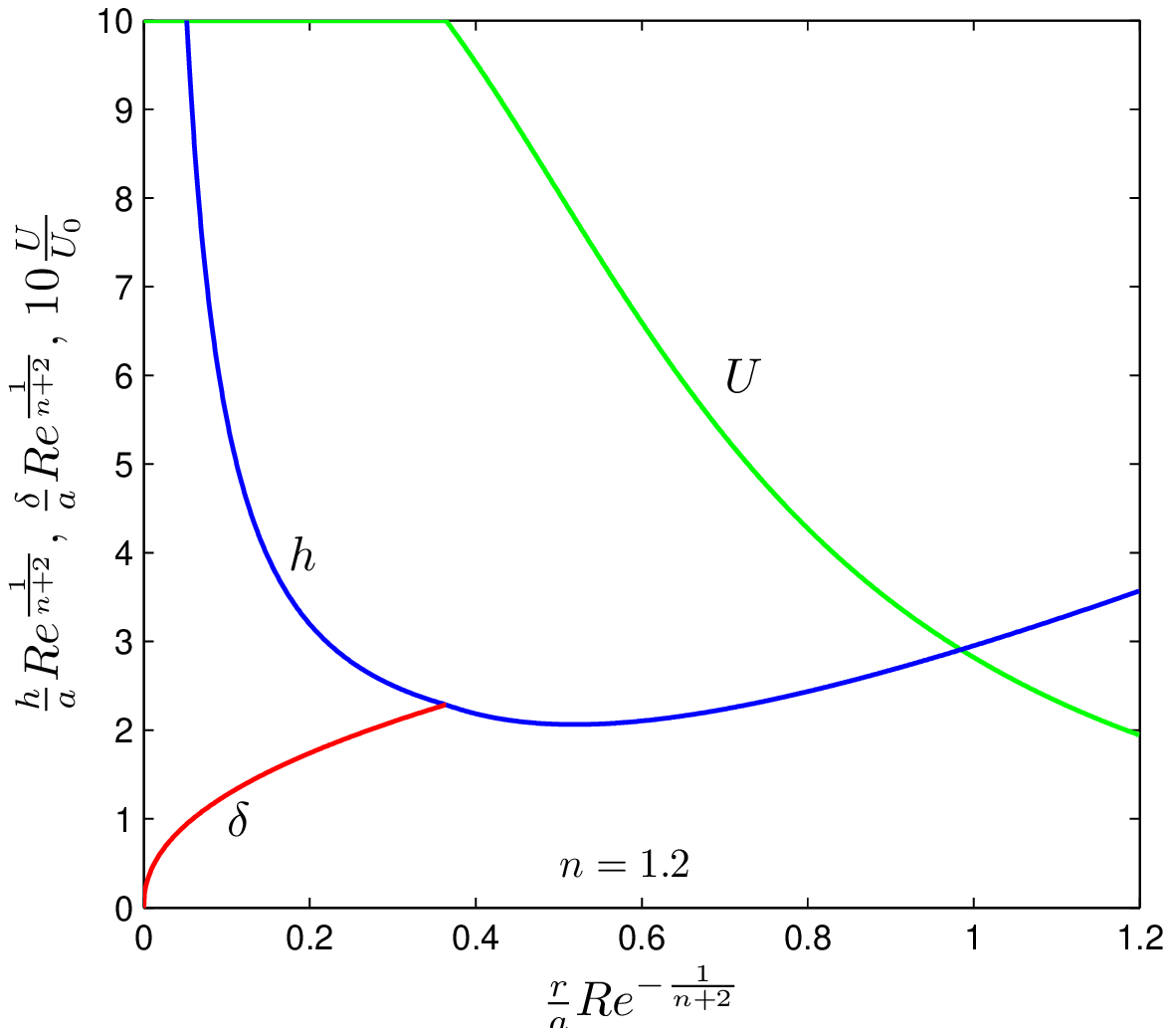}
\end{array}$
\end{center}
\caption{Variation of boundary layer thickness $\delta$, free surface height $h$ and free surface speed $U$ with radial distance, for the flow behavior indices $n = 0.8, 1.0, 1.2$.
Propagation of viscous effect to the free surface, free surface height decay and momentum loss in the flow is delayed with
increase in flow behavior index $n$. Results of Watson are retrieved for $n=1$.   }
\label{fig:contoursStat}
\end{figure}

On applying constant volume flux condition $Q = 2\pi r\Big[U_0 \delta \int_{0}^{1}f(\eta)d\eta+U_0(h-\delta)\Big]$,
free surface height in the region $r \leq r_{b}$ can be obtained as:
\begin{equation}
\label{eq10}
h(r)=\frac{a^2}{2r}+(1-A)\Big[\frac{B(n+1)}{(n+2)}\Big]^{\frac{1}{n+1}}r^{\frac{1}{n+1}}.
\end{equation}

The above expressions (\ref{eq4}), (\ref{eq5}), (\ref{eq7}), (\ref{eq10}), for free surface stream velocity $U(r)$,
free surface height $h(r)$ and the boundary-layer thickness $\delta(r)$ in non-Newtonian
power-law fluids, which leads to the prediction of jump radius, are some significant results of our paper.
Fig. 3 shows the variation of $\frac{U(r)}{U_{0}}$, $\frac{h(r)}{a}Re^{\frac{1}{n+2}}$ and $\frac{\delta(r)}{a}Re^{\frac{1}{n+2}}$,
as functions of $\frac{r}{a}Re^{\frac{1}{n+2}}$, for flow behavior indices 0.8, 1.0, 1.2. It can be observed that,
propagation of viscous effect to the free surface, free surface height decay and momentum loss in the flow is delayed with
increase in flow behavior index $n$. Due to the viscous effect fluid within the boundary-layer loses its momentum as it proceeds upstream,
so the shear-rate decreases with increasing $r$. In a shear-thinning fluid viz. $n=0.8$ effective viscosity increases as the flow progresses,
viscous effect reaches the free surface earlier. If experiments are done fixing the jet Reynolds number jump radius is expected
to be smaller then in a Newtonian fluid ($n=1$) . Opposite behavior is expected for shear-thickening fluid in comparison to
the Newtonian counterpart.

For deriving jump radius, Watson \cite{Watson} applied momentum theorem at the jump, by equating the difference in pressure force across the jump
to the difference in radial momentum flux. Bush and Aristoff \cite{Bush} showed that for circular jumps a radial force due to the
surface tension effects arises at the jump. They considered the geometry of the jump and derived a curvature force
$F_{c} = -2\pi \sigma \Delta H$, where $\sigma$ is coefficient of surface tension and $\Delta H = H - h(r_{j})$.
 By including this correction into the momentum equation they modified the result of Watson as,
\begin{equation}
\label{eq11}
\frac{r_{j}gH^{2}a^{2}}{Q^{2}}\left(1 + \frac{2}{B_{0}}\right) + \frac{a^{2}}{2\pi ^{2}r_{j}H} =
\frac{2r_{j}a^{2}}{Q^{2}}\int_{0}^{h(r_{j})}u^{2} dz,
\end{equation}
where $B_{0} = \frac{\rho gr_{j}\Delta H}{\sigma}$ is the jump Bond number.

As the jump may occur at any point in the development of the flow.
For a power-law fluid we obtain that:
If the jump occurs after $r_{b}$ i.e., $r_{j} \geq r_{b}$,
\begin{equation}
\label{eq12}
\int_{0}^{h(r_{j})}u^{2} dz = \frac{n+1}{3nc}U(r_{j})^{2}h(r_{j}).
\end{equation}
for the jumps occurring before $r_{b}$ i.e., $r_{j} \leq r_{b}$,
\begin{equation}
\label{eq13}
\int_{0}^{h(r_{j})}u^{2} dz = \left[h(r_{j}) + \left(\frac{n+1}{3nc}-1\right)\delta(r_{j})\right]U_{0}^{2}.
\end{equation}
So we get an implicit algebraic expression for the jump radius $r_{j}$
on substituting equations (\ref{eq12}) and (\ref{eq13}) in (\ref{eq11}). Here the flow rate $Q$, jet radius $a$,
jump height $H$, coefficient of surface tension $\sigma$, flow consistency index $K$ and flow behavior
index $n$ are known physical quantities.

In summary, we have studied laminar CHJs in non-Newtonian power-law fluids. Assuming a continues variation in the flow
behavior from Newtonian to non-Newtonian fluids, we have generalized the results of Watson \cite{Watson} for Newtonian fluids,
to generalized-Newtonian fluids obeying the power-law model. Expression for the jump radius is obtained,
which gives new predictions for power-law fluids. Our result reduces to that of Watson for flow behavior index $n=1$.
Surface tension correction term introduced by Bush and Aristoff \cite{Bush}, which becomes significant
for jumps of small radius and height has been taken care of.

We would like to acknowledge Indian Statistical Institute, Kolkata, and Indian Institute of Science Education and Research, Kolkata,
for the academic and financial supports during this work.
One of us (AR) would also like to thank Prof. Prasanta K. Panigrahi for his valuable suggestions and careful reading of
the manuscript, and Sayantan Ghosh for his technical support while writing the manuscript.


\begin{thebibliography}{9}

\bibitem{Mani}
  M. Mathur, R. DasGupta, N.R. Selvi, N.S. John, G.U. Kulkarni, and R. Govindarajan, Phys. Rev. Lett., \textbf{98}, 164502 (2007).
\bibitem{Volo}
  G.E. Volovik, e-print: arXiv: physics/0508215; S.B. Singha, J.K. Bhattacharjee, A.K. Ray, Eur. Phys. J. B, \textbf{48}, 417 (2005).

\bibitem{Rayleigh}
  Lord Rayleigh, Proc. R. Soc. London A, \textbf{90}, 324 (1914).

\bibitem{Watson}
  E.J. Watson 1964, J. Fluid Mech., \textbf{20}, 481 (1964).

\bibitem{Olson}
  R. Olson and E. Turkdogan, Nature, \textbf{211} 813 (1966).

\bibitem{Craik}
  A. Craik, R. Latham, M. Fawkes, and P. Gibbon, J. Fluid Mech., \textbf{112}, 347 (1981).

\bibitem{Bohr1}
  T. Bohr, P. Dimon, and V. Putkaradze, J. Fluid Mech., \textbf{254}, 635 (1993).

\bibitem{Ell1}
  C. Ellegaard, A. Hansen, A. Haaning, and T. Bohr, Physica Scripta, \textbf{T67}, 105 (1996).


\bibitem{Schlichting}
  H. Schlichting, \emph{ Boundary Layer Theory} (McGraw-Hill, New York, 1968).


\bibitem{Liu}
  X. Liu and J. Lienhard, Exps. Fluids, \textbf{15}, 108 (1993).

\bibitem{Ell2}
 C. Ellegaard, A. Hansen, A. Haaning, A. Marcusson, T. Bohr, T. Hansen, and S. Watanabe, Nature, \textbf{392}, 767 (1998).

\bibitem{Yokoi}
  K. Yokoi,and F. Xiao, Phys. Rev. E, \textbf{61}, 1016 (2000).

\bibitem{Bush}
  J. Bush and J. Aristoff, J. Fluid Mech., \textbf{489}, 229 (2003).

\bibitem{Bohr2}
  T. Bohr, V. Putkaradze, and S. Watanabe, Phys. Rev. Lett., \textbf{79}, 1038 (1997).

\bibitem{Hig}
  F. Higuera, J. Fluid Mech., \textbf{274}, 69 (1994); F. Higuera, Phys. fluids, \textbf{9}, 1476 (1997).

\bibitem{James}
  P.D. James and P.P. Darbowaski, Proc. R. Soc. Lond. A, \textbf{460}, 3143 (2004);
D. Filipussi, J. Gratton and F. Minotti, Nuovo Cim., \textbf{116}, 393 (2001).

\bibitem{Bird}
  R.B. Bird, R.C. Armstrong and O. Hassanger, \emph{Dynamics of polymaric liquids}, Wiley (1977).

%\bibitem{Higuera2}
 % F. Higuera, Phys. fluids, \textbf{9}, 1476 (1997).


%\bibitem{Brown}
 % S.N. Brown and K. Stewartson, J. Fluid Mech., \textbf{23}, 673 (1965).

%\bibitem{Fili}
 % D. Filipussi, J. Gratton and F. Minotti, Nuovo Cim., \textbf{116}, 393 (2001).

%\bibitem{Arnab}
 % A.K. Ray, J.K. Bhattacharjee, Phys. Lett. A, \textbf{371}, 241 (2007).

%\bibitem{JKB}
 % S.B. Singha, J.K. Bhattacharjee, A.K. Ray, Eur. Phys. J. B, \textbf{48}, 417 (2005).

%\bibitem{Dabrowaski}
 % Dabrowaski, P.P. 2004 Boundary-layer flows in non-Newtonian fluids.
  %\emph{PhD thesis}, The university of Adelaide, Australia.

 %\bibitem{Berezin}
  %Berezin, Y. A., Chugunov, V. A., Hutter, K.  2001 Hydraulic jumps on shallow layers of non-Newtonian fluids.
  %\emph{J. Non-Newton Fluid Mech.}, \textbf{101}, 139-148.

%\bibitem{Metzner}
 % A. B. Metzener, \emph{ Advances in Chemical Engineering, Vol. 1 } (Acadamic Press, New York, 1956)


%\bibitem{Tani}
 % Tani, I. 1949 Water jump in boundary layer.
  %\emph{J. Phys. soc. Japan },\textbf{4}, 212-215.

 % \bibitem{Brechet}
  %Brechet, Y. and Neda, Z. 1999 On the circular hydraulic jump.
  %\emph{Am. J. Phys. },\textbf{23}, 8.

%\bibitem{Errico}
 % Errico, M. 1986, A study of interaction of liquid jets with solid surfaces.
  %\emph{PhD thesis}, University of California, San Diego.

%\bibitem{Vasista}
 % Vasista, V. 1989, Experimental study of the hydrodynamics of an impinging liquid jet.
  %\emph{B. Eng. thesis}, MIT.


\end{thebibliography}
\end{document}